# Apollonian Packing in Polydisperse Emulsions

Sylvie Kwok,*[a] Robert Botet,[b] Lewis Sharpnack,[c] and Bernard Cabane[a]

We have discovered the existence of polydisperse High Internal-Phase-Ratio Emulsions (HIPE) in which the internal-phase droplets, present at 95% volume fraction, remain spherical and organise themselves according to Apollonian packing rules. These polydisperse HIPEs are formed by emulsifying oil dropwise in a surfactant-poor aqueous continuous phase. After stirring has ceased, their droplet size distributions begin to evolve spontaneously and continuously through coalescence towards well-defined power laws with the Apollonian exponent. Small-Angle X-Ray Scattering performed on aged HIPEs demonstrate that the droplet packing structure agrees with that of a numerically simulated Random Apollonian Packing. We argue that when such concentrated emulsions are allowed to evolve, the coalescing droplets must obey volume and sphericity conservation. This leads to a mechanism that differs from typical coalescence in dilute emulsions.

High Internal-Phase-Ratio Emulsions (HIPE)[1–4] are coarse, long-lasting mixtures of non-miscible liquids in which the internal-phase droplets (e.g. oil) occupy a volume fraction higher than 74%, and are separated by an outer continuous phase (e.g. water + surfactant). These extremely concentrated emulsions were described over a century ago[5], and have since been used industrially for safety and rocket fuel, oil recovery fracturing fluids, an latex foam production.[6–11]

Until now, HIPEs have been made with a very high concentration of surfactant in the continuous phase (7 – 30%, if the surfactant is non-ionic).[1,12,13] Under such conditions, it is relatively easy, using processes inspired by the classical mayonnaise recipe[14], to obtain HIPE in which the inner phase droplets are nearly monodisperse in their diameters.[1,4,15] As their volume fraction increases, the inner-phase droplets remain separated by continuous films of the outer phase, but they become non-spherical with beautiful polyhedral shapes such as rhomboedra, dodecahedra and tetrakaidecahedra packed with short-range order.[1,4,16] The surface tension of the interconnected surfactant films causes these HIPEs to behave like elastic solids, since any displacement of a droplet requires the stretching of this film network. Mason et al. likewise observed the same rheological / mechanical behaviour through the clever use of polymer dialysis to impose very high osmotic pressure on a HIPE in order to attain 97%.[17,18]

In the present work, we show that another class of HIPEs exists, which have completely different structures and properties. They are obtained when surfactant availability is reduced. Indeed, in the experiments reported here, the aqueous concentration of surfactant in the continuous phase was only 0.6 %wt. Surprisingly, these HIPEs flow under the effect of very weak forces such as gravity. When swollen by excess continuous phase, they simply become regular oil-in-water emulsions, characterized by a power-law droplet-size distribution. Such extreme polydispersity causes the structure of the HIPEs to exhibit scale invariance instead of the short-range order in monodisperse emulsions. According to X-ray scattering, the relative positions of droplets match an Apollonian construction, where the interstices between droplets are occupied by even smaller ones (of the largest size possible). This is particularly curious because it has long been believed in the domain of concentrated colloidal science that "this peculiar kind of heterodispersity will rarely, if ever, be encountered"[19] due to the need to fabricate many discrete populations each containing a very large number of droplets of different sizes.

Each HIPE (10 – 20mL in volume) was made with the non-ionic surfactant, hexaethylene glycol monododecyl ether ($C_{12}E_6$), water and a heavy mineral oil. $C_{12}E_6$ was diluted in Millipore Milli-Q water to 0.6%wt to constitute the HIPE's continuous phase. The critical micellar concentration of $C_{12}E_6$ was 0.004%wt under the experimental conditions.[20] Oil was then added dropwise, about 40 – 50μL per drop, under constant shearing from a 4-bladed propeller stirrer at speeds between 200 – 1000rpm. To avoid emulsion inversion, a new oil drop was introduced only after the previous one had been homogeneously dispersed. This protocol may be generalized to other chemical systems, as long as the following conditions are respected: slow addition of the internal phase to avoid emulsion inversion[1,12,14,15]; selection of an appropriate surfactant in accordance to the Bancroft rule[1]; and a sufficiently small quantity of surfactant to allow for some coalescence to occur, which is the novelty that we have introduced in our work.

**Optical microscopy.** Figure 1 presents images of three HIPEs at the same internal phase ratio φ = 0.95 – the oil droplets occupy 95% of the total volume, and the continuous aqueous phase only 5% – with decreasing amounts of surfactant. The left panel shows a classical HIPE in which the continuous phase contains 20% surfactant; the packing is an assembly of distorted micrometer-sized polyhedral oil droplets. In the central panel, the surfactant concentration in the continuous phase is reduced to 5%; the assembly has a hybrid texture in which large nearly spherical drops are dispersed in an array of much smaller polyhedral droplets. The right panel shows a HIPE in which the surfactant is reduced to 0.6% in the continuous phase; all the droplets are perfectly spherical with no signs of being distorted. We noted that droplet deformation occurred only when the amount of $C_{12}E_6$ exceeded 3%wt; we found the same spherical droplet geometry for a HIPE prepared with 2.4%$C_{12}E_6$ at φ = 0.9956.

*sylvie.kwok@espci.fr
[a] Laboratoire Colloïdes et Matériaux Divisés – Chemistry, Biology, Innovation (CBI) UMR 8231, ESPCI Paris, CNRS, PSL*Research University, 10 rue Vauquelin, 75005 Paris, France.
[b] Laboratoire Physique des Solides, CNRS UMR 8502, Univ. Paris-Saclay,, Univ. Paris-Sud, 91405 Orsay, France
[c] ESRF – The European Synchrotron Radiation Facility, CS40220, 38043 Grenoble Cedex 9, France

The change in droplet geometry from spheres to polyhedrons appears to be a gradual continuous transition as function of increasing surfactant concentration.

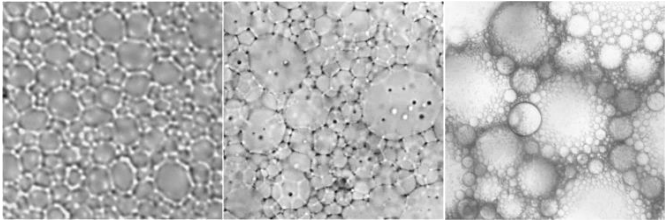

Fig. 1 Optical microscope photos of HIPEs at large volume fractions (ϕ ≈ 0.9). Depending on the surfactant concentration in the continuous phase, the oil droplets have different shapes: polyhedral at 20% surfactant (left); polyhedral (small) and round (large) at 5% surfactant (middle); spherical at 0.6% surfactant (right). The size of each photo is 20μm, 50μm, and 250μm respectively. Droplets < 1μm could not be resolved under optical microscope.

**Droplet-size distributions**. We subsequently focused on the HIPE made at the lowest surfactant concentration ("surfactant-poor", Figure 1, right panel). To quantify the droplet-size distribution, we diluted the HIPE in excess continuous phase and used Dynamic Light Scattering. Fig. 2 shows that the distribution, $n(a)$, is well fitted by a power law within a defined range of diameters:

$$n(a) \propto 1/a^{d_f+1}, \qquad (1)$$

where $a$ is the oil droplet diameter.[21–23] From the exponent of the power law, we may deduce $d_f$, the fractal dimension characterizing the interfaces formed by the droplets.

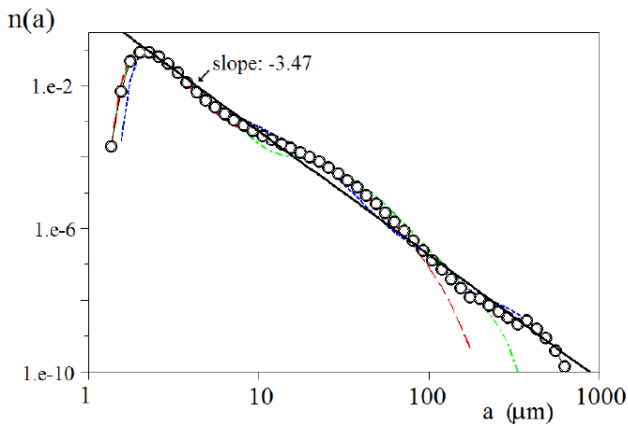

Fig. 2 (colour online) Droplet-diameter distribution in a surfactant-poor HIPE at ϕ = 0.95, measured by Dynamic Light Scattering (Malvern Mastersizer 3000), expressed in double-logarithmic scales to highlight any power-law behaviour. The plotted slope of $\log n(a)$ vs. $\log a$ gives relation (1). The solid line represents a theoretical slope of -3.47 (corresponding to $d_f$= 2.47) and is drawn as a guide. Three samples were prepared under different shearing conditions: at 200rpm, 500rpm and 1000rpm (blue dots, green dots-dashes, red dashes respectively). The open circles represent the values of the droplet-diameter distribution averaged over the three samples.

Regardless of their initial state (as given by different shearing conditions during emulsification), these surfactant-poor HIPEs acquired a power-law diameter-distribution with $d_f$ = 2.48 – 2.50, after they had been left to evolve through coalescence for a month after cessation of stirring, at ambient temperature and at rest. Ostwald ripening is negligible here, due to the heavy mineral oil's insolubility in water. The exponents that we found are particularly interesting because it is known that power-law distributions of spheres can fill space entirely if their exponents have certain values.[22,24] In particular, variants of the Random Apollonian Packing model have been reported to lead to the same diameter distribution (1) with comparable values calculated for $d_f$ (2.45 – 2.52).[25–28]

**Random Apollonian packing.** The process of filling all available space with only spheres of varying sizes is named after Apollonius of Perga, a Greek geometer who famously solved 2200 years ago the problem of finding all cotangent circles to three pre-existing circles.[29] The problem was later generalized by iterating the procedure ad infinitum, and it was found that such iterations would yield a space-filling system of infinitely polydisperse spheres. The Random Apollonian geometrical construction begins with an initial system of a few randomly dispersed spheres, and space is progressively filled as new spheres (the biggest possible at each step) are successively found and added one after another to occupy remaining voids in the packing. In the field of numerical simulation, several different algorithms exist to maximize the size of these newly added spheres at each iteration step.[25–28,30] The resulting size-distribution for the densest packing of these spheres expresses as a power law (1) over an ever-increasing range of diameters ($a_{min}$, $a_{max}$).[24] In the limit of an infinite number of iterations, the interface formed by the spheres is eventually fractal in nature[22] and characterized by the fractal dimension, $d_f$.

To model the spatial arrangement of oil droplets in our HIPEs, we numerically simulated a dispersion possessing the same droplet-diameter distribution (therefore the same $d_f$ and the same ratio $a_{min}/a_{max}$ between the smallest and the largest diameter) at inner-phase volume fraction ϕ (Fig. 3). We validated the relevance of this structure by comparing the Small Angle X-ray Scattering (SAXS) spectra for this model with the HIPEs' experimentally measured SAXS spectra.

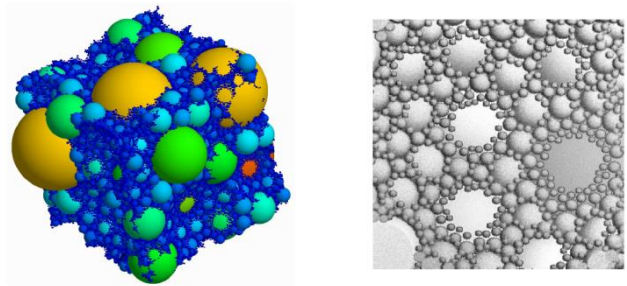

Fig. 3 (colour online) (left) A disordered Apollonian packing of 32000 spheres, ϕ = 0.92, in a cubic box with periodic boundary conditions, generated by the ORAP algorithm. Colours represent the sphere sizes, from red for the largest spheres to blue for the smallest ones; (right) Part of a sliced representation of the same system (in shades of grey), to compare with the experimental image shown in the Fig. 1, right.

**Details of the numerical simulation**. We used a 3D Osculatory Random Apollonian Packing (ORAP) process as a generic algorithm to construct disordered Apollonian packing of spheres. In an ORAP, a location is chosen randomly inside the voids in the packing. A candidate sphere is then centred upon this point such that it is tangent to the neighbouring spheres constituting the void. While there exists several other approaches in the field of numerical simulations to build a Random Apollonian Packing – such as the Packing-Limited Growth model where seeded spheres are immobile and grow until they touch a neighbour[30] – we chose an algorithm based on that described by Varrato and Foffi[28], where a number of positions in the local voids that may be occupied by the newly-added sphere are explored, until said sphere has achieved its largest possible diameter without overlaps with its neighbours. This choice ensured that our simulations were more physically realistic in the context of emulsions where liquid colloidal droplets may undergo displacements in order to find optimal positions for retaining their spherical shape.

We ran our ORAP algorithm many times and consistently found $d_f$ = 2.47, in agreement with $d_f$ values found from other algorithms: 2.474 by the local optimization approach[28], 2.48-2.52 given by a genetic algorithm that recursively solves the global optimization condition in an Apollonian construction[27], and 2.45 by the Voronoi-Delaunay method[26].

**Relevance of the ORAP model: comparison of structure factors.** We validated the selection of the numerical model by comparing its structure factor to that measured experimentally of surfactant-poor HIPEs. For the latter, we first measured $I(q)$, the X-ray intensity scattered by $N$ droplets in a small volume $V$ of the HIPE. We also measured $P_{exp}(q)$, the average form factor of these $N$ individual droplets, taking into account their size distribution. Then, according to the classic equation as proposed by Vrij[31] and accepted by nearly all authors in the field of concentrated dispersions and emulsions, comparing $I(q)$ to $P_{exp}(q)$ yields the experimental structure factor, $S_{exp}(q)$:

$$I(q) = \frac{N}{V} P_{exp}(q) S_{exp}(q) \qquad (2)$$

The SAXS experiment was performed on the ID02 instrument at European Synchrotron Radiation Facility.[32] Fig. 4 compares $S_{exp}(q)$ of our three HIPEs at ϕ = 0.95, and the calculated structure factor, $S_{sim}(q)$, of a numerically simulated population of spheres packed according to the ORAP algorithm at ϕ = 0.92. This particular value for ϕ was the highest that we could attain; beyond 0.92, the number of particles necessary for the simulation and computing time required become unacceptably high. We checked numerically for 0.84 ≤ ϕ ≤ 0.92 that the main peak of $S_{sim}(q)$ was almost insensitive to the actual value of ϕ; indeed, an increase in volume fraction generates a large number of very small particles whose contributions modify $S_{sim}(q)$ only in the range of larger $q$ values, far from the main peak being examined. We found that our experimental and calculated $S(q)$ shared the same distinctive features: a depression at low $q$ values, followed by a steep rise to a broad peak of magnitude $S_{max}$ = 1.1 – 1.2, and a tail that rapidly approaches unity at higher $q$ values.

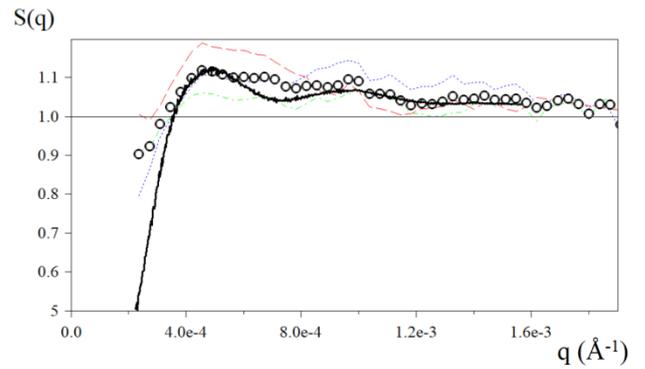

Fig. 4 (colour online) Summary of $S_{exp}(q)$ measured by SAXS for three power-law polydisperse HIPEs at ϕ = 0.95, prepared at 200rpm, 500rpm and 1000rpm (blue dotted, green dot-dashed, red dashed lines respectively), aged 1 month. The open circles represent the average value of the three $S_{exp}(q)$. The heavy solid line is $S_{sim}(q)$ calculated from a simulated ORAP at ϕ = 0.92, displayed in Fig. 3. For the latter, the horizontal axis has been rescaled by by the Wigner-Seitz radius $(3/4\pi n)^{1/3}$, where $n$ is the number of spheres per unit volume.

Among these features, the most distinctive is the small value of $S_{max}$, which is consistent with a system without any sign of a translational order. This is particularly evident in the case of $S_{sim}(q)$, since an ORAP generates systems with scale-invariance and no translational order. Indeed, our $S_{max}$ values are significantly lower than those of colloidal systems with the same ϕ but exhibiting translational short-range order. A striking counterexample is that polymer films produced through the coalescence of monodisperse latex particles, in which $S_{max}$ rises above 2.0 when the films are dried and annealed.[33–35] Thus, the quantitative agreement between our $S_{exp}(q)$ and $S_{sim}(q)$ is a strong indication that in the case of low surfactant availability in the continuous phase, oil droplets in polydisperse HIPE formulations arrange themselves into an ORAP-like structure. We caution here that this result does not mean that our surfactant-poor HIPEs evolved through the same mechanism as an ORAP: in particular, ORAP does not allow for growth or fragmentation of the spheres, whereas oil droplets in surfactant-poor HIPEs may change sizes and volumes by swapping oil and surfactant.

**Coalescence in Apollonian HIPEs**. In fact, coalescence is one of the origins of Random Apollonian Packing in our surfactant-poor HIPEs. This role of coalescence has been reported by Aste[36], who deposited metallic vapour onto a substrate and observed the fusion of the condensed drops giving rise to an Apollonian packing. We argue, in addition, that the particular characteristics of our emulsions, with such high ϕ and such poor surfactant availability, presents a unique set of geometrical constraints that govern the evolution at rest of these spherical droplets: they have to find configurations satisfying the criteria of conserving volume and sphericity. The latter condition is necessary to avoid an unfavourable increase in elastic energy due to droplet deformation. Then, every successful coalescence event between two droplets in such a crowded emulsion may produce multiple spherical daughter droplets that maximally fill the principal void left behind, as well as any neighbouring voids. Such a construction, where radii are locally maximized without overlaps, is consistent with the definition of an Apollonian construction[22]. Since the resulting number of droplets after a coalescence event would increase, we named our proposed mechanism "coalescence-fragmentation". This process leads to a power-law diameter-distribution of spherical droplets that optimally fills space by lodging the right sizes at the right places, i.e. a global Random Apollonian Packing. In our experiments, the Porod domain in our collected SAXS data reveals that the specific surface of the systems increases regularly with time. This observation is in favour of the coalescence-fragmentation mechanism as the driving force of evolution in these systems.

Our calculations show that two emulsion droplets fusing into one – typical coalescence in dilute systems – fails to give a power-law diameter-distribution and rapidly leads to macroscopic phase separation. We used numerical simulation to model our coalescence-fragmentation mechanism in highly concentrated systems of spheres at comparable volume fractions to our surfactant-poor HIPEs. By allowing pairs of coalescing spheres to fission into multiple non-overlapping daughter spheres, our Monte-Carlo simulations demonstrated that a power-law distribution of diameters was indeed obtained at the end of the process with an Apollonian exponent. It appears that a mechanism where coalescence and fragmentation occur in tandem is a novel feature of the evolution in concentrated emulsions: until now, coalescence has never been permitted in a HIPE due to the abundant amounts of surfactants with which they have so far been formulated. Moreover, these two processes have always been considered as independent and of opposing consequences, with one suppressed or ignored to simplify the study of the other in colloidal sciences.

## Conclusions

In conclusion, we have discovered a simple, highly reproducible experimental protocol to fabricate emulsions that spontaneously pack their constituent droplets as a scale-invariant Apollonian liquid. Through elementary steps (coalescence and fragmentation) obeying geometrical rules, adjacent spheres swap volume and rearrange themselves without overlaps, ultimately leading to a structure composed of a power-law polydisperse population of droplets. Consequently, the process attains much higher internal-phase ratios (ϕ can reach at least 0.95) than translational repetition of a single type of spherical droplet.

We believe that our novel approach of permitting coalescence in such concentrated emulsions, through drastically reducing surfactant availability, may be of fundamental interest to physicists and chemists alike: our experimental observations, to be reported in a subsequent publication, suggest that coalescence in such a crowded structure significantly differs from the typical coalescence mechanism known in dilute emulsions.

A Random Apollonian Packing of emulsion droplets may also be of practical use in materials science. Emulsion-templating, an existing application of HIPEs, could be employed to create ultra-dense or ultra-porous solids. The fact that Apollonian emulsions possess oil-water interfaces that are fractal in nature is also a non-trivial property in conductivity-related applications: many theoretical models on thermal and electrical conductivities have been developed for fractal structures, but empirical verification of these models have been largely restricted to dendritic structures[37] or naturally-occurring porous rocks[38]. Thus, our very generalizable protocol for preparing Apollonian emulsions may prove useful by way of obtaining fractal structures of known internal volume fraction and measurable fractal dimension, contributing to the empirical study of transport phenomena in intricate structures.

## Acknowledgements

We wish to express our sincere gratitude towards Yeshayahu Talmon, Lucy Liberman and Irina Davidovich of Technion Israel Institute of Technology, and Klaus Eyer of ETH Zurich for their indispensable help with microscopy. We would also like to thank beamline staff on ID02 at ESRF, Theyencheri Narayanan and Michael Sztucki, for their invaluable technical assistance.